\documentstyle[12pt]{article}

\begin{document}
\title{The New Universe\\
Fixed by a Standing Wave Particle Model}

\author{Rafael A. Vera\thanks{email: rvera@buho.dpi.udec.cl}\\
Deptartamento de Fisica\\
Universidad de Concepcion\\
Casilla 4009. Concepcion. Chile}

\date{Sept 8 1995}
\maketitle

\vspace{-9cm}

\hfill astro-ph/9509053

\vspace{9cm}

\begin{abstract}
The theoretical properties of the black holes (BHs) and of the universe were
derived from a unified relativistic theory based on a generalization of local
relativity for nonlocal cases in gravitational fields and a quantized standing
wave particle model that accounts for relativity, quantum mechanics and the
gravitational tests (See gr-qc/9509014). They fix an isentropic
and conservative steady state that is independent on an eventual universe
expansion because matter also expands itself in the same proportion. The new
black holes BHs resulting from linear properties of the model, after capturing
enough radiation, would explode. Statistically, matter would evolve,
indefinitely, in rather closed cycles between gas and BH states, and
vice-versa. The expected astronomical objects and cosmic radiation backgrounds
that are consistent with the observed facts. This leads to non conventional
models for some celestial objects.
\end{abstract}

\newpage

\section {Introduction}

The main purpose of this work is to find the cosmological and
astrophysical context fixed by {\it a nonlocal (NL) relativistic theory
 based on a single postulate on the common nature of matter and
 stationary forms of radiation's}. In this way the new unified contexts
of all of them: physics, astrophysics and cosmology, would be fixed,
ultimately, by properties of radiation's.

According to previous works \cite{V77}, \cite{V81a}, \cite{V81b},
 \cite{V86}, \cite{V95a}), \cite{V95b}, the simplest particle model that
can account for the general properties of matter and of it's
gravitational (G) field, is
a {\it standing wave} (SW) model. This one is made up of a
quantum of radiation in some stationary state of a well-defined local
frequency ($\nu$) fixed by the space properties.

According to this model {\it all the
local bodies in a system, including the instruments or any standing
wave, must have the same kind of relativistic changes, in the same
proportions, under the same changes of velocity and G field
potential}. Observe that only In this way the local relative values can
always remain constant, i.e., local relativity can be strictly true.

In these works it is also proven that {\it local relativity} is not well
defined for relating quantities measured in different G field
potentials because, according to G time dilation, the clocks and the
atoms of observers in different field potentials are not striclty the same
relative to each other. This means that, in general,{\it the measurement
units} of observers in different field potentials {\it are different
relative to each other}. Then to get strictly
homogeneous relations between quantities measured in different G
field potentials, each of them must be reduced to {\it a common unit
system}.

These quantities, referred to a standard that has not had the same
changes as the objects, are called {\it nonlocal} (NL){\it  quantities}.
They correspond to a generalization of those of {\it local relativity}.
They can be functions on both $\beta =v/c$ and on the NL G field
potentials of the object and of the standard. The last one, whose
position is fixed, can be stated by a sub index. This fixes an
invariable (flat) reference framework.

The model NL relativistic and quantum mechanical properties come
out from the constructive interference of its {\it wavelets}. On the
other hand the NL properties of its long range {\it G field} turn out
to depend only on {\it the relative perturbation rate of the space, $w(r)$,
produced by destructive interferences of random (or out of) phase
wavelets}. Since the net wave amplitude is null, then there is no field
energy. This means that static G fields do not exchange energy with
free bodies or radiation's. The same result comes out from other
ways \cite{V81a}\cite{V81b} \cite{V95b}. This contradicts rather
conventional concepts used by Einstein  in this theory of General
Relativity \cite {E55}.

On the other hand it has been found that, in order that some
gradient of $w(r)$ can exist, some kind {\it wavelet red shift} (WRS)
proportional to the NL distances should also exist. This one is
obviously consistent with the {\it Hubble red shift} (HRS) of light. If
$r^{ik}$ is the NL distance between the positions $i$ and $k$, and if
R is the typical NL distance at which the wavelets are attenuated by
the factor $e^{-1}$,  then the total relative perturbation rate of the
space at a point $i$, compared with that coming from a universe of
uniform density, is:
\begin{equation}
\label{1.1}
w(i) = \sum_k^\infty G\frac{h\nu (r^k)\exp (r^{ik}/R)\,}{r^{ik}}
=\sum_j^\infty G\frac{m(r^j)\exp (r^{ik}/R)\,}{r^{ij}}
\end{equation}

$R$ corresponds with the Hubble radius.

For a universe of uniform density, $w(U) = 1$.

On the other hand the NL gradients in this field turn out to be related
by:
\begin{equation}
\label{1.2}
\frac{\nabla \nu _{r^{\prime }}(0,r)}{\nu _{r^{\prime
}}(0,r)}=\frac{\nabla m_{r^{\prime
}}(0,r)}{m_{r^{\prime }}(0,r)}=\frac{\nabla \lambda _{r^{\prime
}}(0,r)}{\lambda
_{r^{\prime }}(0,r)}=\frac{\nabla c_{r^{\prime }}(r)}{2c_{r^{\prime
}}(r)}=\nabla \phi (r)=-\nabla w(r)
\end{equation}

The first ones correspond to the phenomena of G red shift (GRS),
G work, G contraction and G refraction, respectively.  $\phi(r)$ is
called {\it NL field potential}. The first order approximations of the
above relations correspond with those of general relativity (GR).

They account for all of the ordinary {\it G tests} \cite{V81b}.

\section{The new cosmological context}

\subsection{Model expansion versus universe expansion}

It is currently assumed that matter does not expand itself during an
eventual universe expansion. {\it If this were true} some standard
model could be used as the base for a non expanding theoretical
reference framework. In it, ${dr(ik)}/{r(ik)}=Hdt $. Then, from (1.1),
(1.2), and NL mass-energy conservation,
\begin{equation}
\label{2.1}
d\phi (i)=-dw(i)=-\,d\sum_{k=1}^\infty \frac{Gm(k)f(ik)}{r(ik)}%
=\left[ \sum_{k=1}^\infty \frac{Gm(k)f(ik)}{r(ik)}\right] Hdt,
\end{equation}
\begin{equation}
\label{2.2}
d\phi \left( i\right) =w(U)Hdt=Hdt=dr(ik)/r(ik)=d\lambda
(i)/\lambda (i).
\end{equation}

This means that {\it every wave and particle would expand itself in
the same proportion as the intergalactic distances}. An eventual
universe expansion would not change the relative values of all of
them: the distances, the velocities, the temperatures, the WRS, the
HRS, and therefore, the local physical laws\footnote {This
generalizes the relativity postulates for eventual universe
expansions.}. {\it Then the universe would have not a well-defined
age and it may last indefinitely}. Of course, this would invalid the
current deductions normally made for the universe age, which anyway
seem to be not consistent with the last measurements made with the
Hubble telescope.

\subsection{The new kind of black hole (BH)}

The new exponential G relations have not a singularity at $r=2GM$.
Then, the new kind of BH is different to that of GR \cite{V81b}. Its
nucleus would be just a neutron star (NS) with a strong external {\it
gradient of the NL refraction index } that would act as a mirror for
most of the internal radiation's. Its outcoming {\it critical
reflection angle}, given
by $sin^{-1}[\left( 2eGM/r\right) e^{-2GM/r}]$, would be rather
negligible. Thus the escape probabilities would be not strictly null.
Then the BH would absorb and store for long time most of the
radiation's traveling within the impact parameter $2eGM$.

\subsection{Relativistic particle generation}

In a way similar to the earth {\it auroras}, most of the positively
charged nuclei would be driven by the magnetic fields towards the
BH polar regions. Since the neutron binding energy in a BH is of a
higher order of magnitude compared with that in atoms, then one the
most probable reactions between them is {\it nuclear stripping} \cite
{V77}, \cite{V81b}. In it, some of the atomic neutrons
would be captured by the BH while the remaining nucleus (proton or
proton rich nucleus) would be rejected by the NS. The last one
would take away the NL mass-energy difference between the
original and final states of the captured neutrons. They could only
escape from the magnetic fields, in axial directions, within the small
escape angle given above. They would form {\it narrow jets of
relativistic particles} richer in protons and with higher energies for
higher $p/n$ ratios. They are consistent with the composition and
energies of {\it cosmic ray particles} \cite{V81b}, \cite
{R81}. They are also consistent with the {\it radio sources
and jets} going away from central regions of galaxies, most of them
in just the expected orientations.

This process would be most important because it would convert G
work into mechanical and nuclear latent energies. This would
regenerate new gas of high nuclear and kinetic energies at the cost
of rather burnt out materials like He or heavier elements. Such low
entropy materials can in principle extend the luminous lifetimes of
galaxies beyond the limits estimated from the current models.

\subsection{The entropy switch}

{}From the BH surface, just to the contrary of the outside regions, the
external universe would look as a source of {\it blue shifted
radiation's} that would increase both the local temperature and the
probabilities for filling up the local SW levels up to the highest NL
frequencies. This is equivalent to a decrease of the local entropy. In
this way the average NL mass and kinetic energy of the nucleons
would increase with the time, with the radiation energy coming from
the rest of the universe, up to some unstable state in which any
decrease of the NL refraction index gradient generated by external
bodies would produce {\it frustrated reflections} that can trigger the
mass outflow. Thus the BH can {\it explode} producing low density
gas flowing away through older stellar remnants orbiting around it.
This would transform a fraction of the kinetic energy into rotational
one associated with randomly oriented angular momentum's. This is
also consistent with the fronts of H rich matter diverging from very
small regions in the universe.

\section{The new astrophysical context}

{}From above the universe would last, indefinitely, in a kind of
conservative and {\it isentropic steady state}.  In it, {\it matter and
radiation's would evolve, indefinitely, in rather closed cycles,
between the states of gas and BHs, and vice versa}.

\subsection{Matter cycles}

Single and chain of BH explosions would produce {\it rather
spherical stellar clusters and elliptical galaxies,  rather free of
metals}. They would regenerate randomly oriented angular
momentum's that, in the long run, would be canceled out at faster
rates compared with those parallel to the galactic axis. Thus an {\it
elliptical galaxy} would progressively get {\it disc and spiral shapes}
of smaller volumes. Finally it would become reduced to a small
central luminous volume (AGN and {\it quasar}) with massive stars
and high density black bodies (black holes, neutron stars)
surrounded by a halo of dead stars and planetesimals [{\it black
galaxy}]. The explosive events as supernovas would produce large
changes of luminosity, within relatively short periods, that are
consistent with those of quasars \cite {N90}.

Due to the low $\phi (r)$ in the black galaxy center, their atoms
would emit strongly red shifted light rather scattered and reflected by
the external bodies. This accounts for the fact that quasar
correlation's improve under the assumption that most of the
observed red shift is intrinsic \cite {B73}. The detection of
metal lines would also prove the existence of highly evolved (old)
matter.

The {\it black galaxy} (BG) resulting from a luminous one would be
cooled down by its BHs. It would also capture and store radiation
coming from the external universe, in a way similar to a huge BH.
After a long period, the explosion of some central BH can trigger a
chain of BH explosions that would regenerate a luminous galaxy.

Within a larger time scale, the galaxy regeneration would look like a
BG explosion that can trigger the virtual explosions of the next BGs,
and so on. They would produce {\it clusters}. Superclusters would also
be due to similar mechanisms. Thus the {\it fronts} of galaxies in
luminous stages would also account for the large scale structure of
the universe.

\subsection{High energy step down in stellar objects}

Mechanisms of nuclear stripping similar to those occurring BHs
could also occur during the matter fall over neutron stars (NSs),
either steadily or in pulsed ways. They may also occur rather hidden
inside some stars or gas clouds. They would transform heavy (burnt
out) elements into protons of higher kinetic and nuclear latent
energies that would promote convection currents. This would
prevent overheating and stellar collapse after neutrino cooling.

This kind of stellar model, \cite{V93}, is consistent with all of
them: the low neutrino luminosity's, the higher densities and
temperatures, the better defined mass-luminosity's relations and the
magnetic structures of {\it main sequence stars}.

\subsection{Density and isotropy of the universe}

Due to the higher rates of energy emitted by the luminous galaxies
compared with those absorbed by the BGs, it is inferred (after a
mass-energy balance) that {\it most of the universe should be in the
state of low temperature BGs}, cooled down by their own BHs. This
is consistent with the high average density of the universe derived
from (1.1), and assuming $H=75 km/\sec $ per mps\footnote {When
the common mass and energy unit is the joule, $G = G_{newt} c^
4$.}.This one is $ \simeq [4\pi GR^2]^{-1}$, i. e.,  about $ 10^{-29}
gm/cm^3 $. This is of a higher order of magnitude than that of the
luminous fractions of the universe. This is also consistent with the
current mass excesses detected from dynamic methods in galaxies
and clusters.

After integration of (2), the space properties are fixed, mostly, by
matter existing between $R$ and $3R$. The contribution of
relatively local matter is extremely small compared with that of the
rather uniform universe. This is consistent with {\it the weakness of
ordinary G interactions, and with  the high isotropy of both the space
properties and of the cosmological radiation background}.

The low temperature black-body radiation coming from BGs, red
shifted during its long average trip up to the observer, ($2R $),
would fix {\it a rather uniform low temperature cosmic radiation
background}. Thus {\it the universe would always look like a perfect
radiation absorber}\footnote {Only steady state cosmologies can
account for the {\it arrows} in nature \cite {N90}.}.

\section{Conclusions}

The theoretical properties of the SW particle model fix a new kind of
{\it conservative and isentropic steady state} in which matter and
radiation's evolve, indefinitely, in rather closed cycles. These cycles
are fairly consistent with the luminous bodies ranging between
elliptical galaxies and quasars, and also with larger scale structures
of the universe.

This theory opens the way for new stellar models and non
conventional interpretations of many celestial phenomena. The new
universe would have not the narrow limits of time fixed by the rather
conventional theories. In this way, also,  astrophysics could do
without the relatively large number of non testable hypotheses that
can be advanced on the universe origin.

There is simultaneous consistency of the theoretical properties of
the SW model with fundamental physics, and of the new
cosmological context with a wide range of astronomical
observations. This seems to be a fair reliability test for all of them,
the SW particle model, for the relationships derived from it, and for the
new cosmological and astrophysical contexts. This unified way may
contribute to understand nature in terms of the most elemental
properties of radiation's (or vice versa), thus depending on the
minimum number of parameters, postulates, and arbitrary
assumptions normally made on relations between matter and its G field.

Due to the large amount of subjects and materials accumulated
from 1976 up to day, the author intends to complilate all of this
work into a single book for that may be useful to those that may
like to go in this way for undestanding nature from a self-consistent
and unified viewpoint\cite{V95c}.

Acknowledgement. I appreciate very much the help of TW Andrews, after
sending me some helpful literature, including his own ideas. I would also
appreciate some encouragement and colaboration for finding
further astronomical tests for this theory.

\enddocument